\documentclass{article}
\pdfoutput=1
\usepackage{jcappub}
\usepackage{epsfig}
\usepackage{graphicx}
\usepackage[latin1]{inputenc}
\usepackage{amsmath}
\usepackage{url}
\def\simlt{\ \raise -2.truept\hbox{\rlap{\hbox{$\sim$}}\raise5.truept   %
\hbox{$<$}\ }}
\def\simgt{\ \raise -2.truept\hbox{\rlap{\hbox{$\sim$}}\raise5.truept   %
\hbox{$>$}\ }}                                                          %

\def\be{\begin{equation}}
\def\ee{\end{equation}}
\def\newline{\hfil\break}

\def\la{\mathrel{\hbox{\rlap{\hbox{\lower4pt\hbox{$\sim$}}}\hbox{$<$}}}}
\def\ga{\mathrel{\hbox{\rlap{\hbox{\lower4pt\hbox{$\sim$}}}\hbox{$>$}}}}

\title{The impact of non-Planckian effects on cosmological radio background}

\author[a,1]{Sergio Colafrancesco,\note{Corresponding author.}}
\author[a]{Mohammad Shehzad Emritte,}
\author[a]{Paolo Marchegiani}

\affiliation[a]{School of Physics, University of the Witwatersrand, Private Bag 3, WITS-2050, Johannesburg, South Africa}

\emailAdd{Sergio.Colafrancesco@wits.ac.za}
\emailAdd{emrittes@yahoo.com}
\emailAdd{Paolo.Marchegiani@wits.ac.za}

\abstract{Non-Planckian (NP) spectral modifications of the CMB radiation spectrum can be produced due to the existence of a non-zero value of the plasma frequency at the recombination epoch.
We present here an analysis of NP effects on the cosmological radio background and we derive, for the first time, predictions of their amplitude on three different observables: the CMB spectrum, the Sunyaev-Zel'dovich (SZ) effect in cosmic structures, and the 21-cm background temperature brightness change.
We find that NP effect can manifest in the CMB spectrum at $\nu \simlt 400$ MHz as a drastic cut-off in the CMB intensity. Using the available CMB data in the relevant $\nu$ range (i.e., mainly at $\simlt 1$ GHz and in the COBE-FIRAS data frequency range), we derive upper limits on the plasma frequency $\nu_p$ = 206, 346 and 418 MHz at 1, 2 and 3 $\sigma$ confidence level, respectively. 
We find that the difference between the pure Planck spectrum and the one modified by NP effects is of the order of mJy/arcmin$^2$ at $\nu \simlt 0.5$ GHz and it becomes smaller at higher frequencies where it is $\sim 0.1$ mJy/arcmin$^2$ at $\nu \simgt 150$ GHz, thus indicating that the experimental route to probe NP effects in the early universe is to observe the cosmological radio background at very low frequencies.
We have calculated for the first time the NP SZ effect (SZ$_{NP}$) using the upper limits on $\nu_p$ allowed by the CMB data.  We found that the  SZ$_{NP}$ effect shows a unique spectral feature, i.e. a peak located exactly at the plasma frequency $\nu_p$ and this is independent of the cluster parameters (such as its temperature or optical depth). This offers therefore a way to measure directly and unambiguously the plasma frequency at the epoch of recombination by using galaxy clusters in the local universe, thus opening a unique window for the experimental exploration of plasma effects in the early universe. We have shown that the SKA-LOW has the potential to observe such a signal integrating over the central regions of high-temperature clusters.
The studies of NP effects through the SZ$_{NP}$  can be done by intensive observations of only one galaxy cluster, or with a stacked spectrum of a few well known clusters, thus avoiding the need of large statistical studies of source populations or wide area surveys.
Finally, we also show that future low-$\nu$ observations of the cosmological 21-cm background brightness temperature spectral changes have the possibility to set global constraints on NP effects by constraining the spectral variations of the temperature brightness change $\delta T_b$ induced by the plasma frequency value at the epoch of recombination. 
}

\begin{document}
 \maketitle

\section{Introduction}
The existence of a Planck spectrum for a system filled with a black body (BB) radiation assumes the presence of thermodynamic equilibrium between photons and matter.
Interaction between photons and matter must be small enough for the gas of photons to be considered ideal and to avoid substantial absorption and irreversible attenuation of electromagnetic radiation in the system (we note here that the interaction between photons is extremely weak). 
At the same time, the presence of matter and the weak interaction between matter and radiation is necessary for the photon gas to be in equilibrium \cite{LL}.
These conditions are valid for a BB radiation from a photon gas and hence are also expected to be valid for the cosmic microwave background (CMB) spectrum emerging at the epoch of recombination (at redshift $z_{rec} \approx 1400$) because the CMB spectrum is assumed and measured \cite{Fixsen1996} to be close to a Planckian BB spectrum.\\

In this context,  the dispersion relation for electromagnetic radiation in non-ionized media (e.g., a rarefied gases) is given by 
\begin{equation}
\omega = c k
\end{equation}
where $k$ is the wavenumber, $c$ the speed of light and $\omega$ the angular frequency of the photon. However, in a medium which is ionized (e.g., an ionized plasma), the analogous dispersion relation is given by  \cite{Triger2010}
\begin{equation}
\omega^2 = {c^2 k^2 + \Omega_p^2}
\end{equation}
where $\Omega_p =2 \pi \nu_p$ is the angular plasma frequency, with $\nu_p^2=\sum_{i=1}^{N} n_i e^2/4 \pi m_i \epsilon$ being the actual plasma frequency, where the sum is over all charged particles species. The existence of such a dispersion relation is due to the coupling between the electromagnetic waves (photons) and the collective behavior of the plasma \cite{Kittell1986}. Consequently, photons with frequencies lower than the plasma frequency $\nu_p$ do not exist in the final radiation spectrum. The presence of a non-zero value of the plasma frequency (in the following we use for our convenience the a-dimensional plasma frequency $x_p= h \nu_p/k_B T_{CMB}$, with $k_B$ being the Boltzmann constant and $T_{CMB}$ the CMB temperature today) will cause the intensity of the photon spectrum to be zero for frequencies less than $x_p$ and this fact will reflect as a modified blackbody spectrum subject to non-Planckian (NP) effects. A general form for such a NP spectrum has been derived by \cite{Triger2010} and \cite{Medvedev1998} for astrophysical plasma-photon systems, such as the CMB and $\gamma$-ray bursts atmospheres. 
We note here that the plasma frequency $x_p$ depends on the number density of all charged particles in the plasma, and hence deviation from Planck spectrum will also depend on this number density. 

In principle, the plasma frequency can be computed using knowledge of the cosmological recombination history. For example, a simple result obtained using a number density of 300 electrons/cm$^3$ is that the plasma frequency at recombination is $\approx 0.2$ MHz \cite{Triger2010}. However, other effects bring uncertainties on this result. In fact, in addition to the ionized plasma, dark matter annihilation can produce electrons and positrons which will perturb the ionization fraction \cite{Chen2004, Padmanabhan2005, Huetsi2009, Galli2009, Dvorkin2013}. It has been shown that the ionization fraction can increase due this effect by more than an order of magnitude providing a plasma frequency of order of $\approx 1.5$ MHz \cite{Dvorkin2013}. Furthermore, electrons and positrons (hereafter we refer to these particles simply as electrons) can also be produced by hadronic collisions \cite{Dermer1986}, originating another additional source of free particles, and consequently an increase of the plasma frequency value.

The CMB spectrum is also vulnerable to other distortions like the $\mu$ and $y$ distortions, related respectively to a frequency-dependent chemical potential and the Compton scattering of the CMB photons (see \cite{Tashiro2014} for a review and references therein). The plasma effects in principle could also create a distortion well before the recombination era. Since the free particles number density depends on redshift, the plasma frequency for the CMB will also depend on redshift as $\nu_p \propto (1+z)^{3/2}$. Other distortions that can occur in the early universe are the Silk damping of small-scale perturbations, the cooling of photons by electrons and baryons, the decay and annihilation of relic particles, primordial magnetic fields and evaporating primordial black holes (see \cite{ChlubaSunyaev2012, Tashiro2014, Chluba2014} and references therein).

It is well known that some processes like Compton scattering, double Compton scattering and bremsstrahlung can change the photons energies or inject photons in the universe, in principle leading to a thermalization process, causing loss of information about the distortions occurring in the CMB \cite{ChlubaSunyaev2012, Tashiro2014, Procopio2009}. However this does not seem to be possible for distortions due to the plasma frequencies. In fact, the time scales of these processes \cite{Tashiro2014}, $t_{comp} =1.23 \times 10^{29} / (1+z)^4$ s, $t_{double}=1.34 \times 10^{40} [x^3/(e^x -1 )] / (1+z)^5$ s and $t_{free}= 8.59 \times 10^{26} [x^3/(e^x -1 )] / (1+z)^{2.5}$ s, are larger than the plasma time scale defined as $t_p = 1/\nu_p$ \cite{Padmanabhan2000}. For example, at $z=1400$ and for a plasma frequency of 0.2 MHz corresponding to $x_p=2.5 \times 10^{-9}$, we obtain the values $t_{comp} =3.2 \times 10^{16}$ s, $t_{double} = 1.5 \times 10^7$ s, $t_{free}=72$ s and $t_p=6.4 \times 10^{-6}$ s; at $z= 10^6$ and a plasma frequency of 3 GHz corresponding to $x_p=6.64 \times 10^{-8}$, we obtain $t_{comp} =1.2 \times 10^5$ s, $t_{double} = 5.9 \times 10^{-5}$ s, $t_{free}=3.8 \times 10^{-5}$ s and $t_p=3.4 \times 10^{-10}$ s. The role of these processes, is, therefore, to eliminate the effects due to other distortions (see, e.g., \cite{Illarionov75a, Illarionov75b, Danese77, Danese82, Burigana91a, Burigana91b, Hu93a, Hu93b}), and to lead the CMB spectrum at the recombination to have the BB shape modified by the plasma frequency. This is the spectrum we can measure, and from which we can derive our results about the plasma frequency.

So far, only theoretical arguments on NP plasma effects on the CMB have been discussed and neither experimental limit has been derived nor observational strategies to probe these effects has been proposed.
All these uncertainties make it desirable to obtain limits to the plasma frequency from observations, that can be used to constrain the information about the processes occurred during and before the recombination. 

In this paper we want to derive for the first time constraints on the value of $x_p$ from the spectral shape of the CMB using current observations. In order to check if there is evidence for a non-zero value of $x_p$ in the CMB spectrum, we therefore perform a comparison of the theoretical expectation for a NP CMB spectrum with the available observations at both low (radio) and high (microwave and mm.) frequencies. While the knowledge of the CMB spectrum is  very precise in the microwave and mm. frequency regions after the COBE-FIRAS measurements \cite{Fixsen1996}, its knowledge at lower radio frequencies (particularly at frequencies below 1 GHz where the NP effects are expected to emerge) is still quite uncertain \cite{Howell1967,Sironi1990,Sironi1991}. Therefore, in Sect.2 we will investigate first the limits on NP effects on the CMB spectrum and on the value of $x_p$ from the comparison with CMB measurements.

We further stress that the modification on the CMB spectrum at low-$\nu$ has an effect on other observable quantities that depend on the CMB spectrum.  In this paper we will consider two additional cosmological observables whose spectral shapes are modified by the presence of NP effects in the CMB spectrum: i) 
the Sunyaev-Zel'dovich (SZ) effect produced by the inverse Compton scattering (ICS) of CMB photons off the electrons residing in the atmospheres of cosmic structures like galaxy clusters, radio galaxies and galactic halos \cite{Colafrancesco2014}, and ii) the temperature brightness change of the cosmological 21-cm background as produced by the evolution and reionization history of the primordial plasma during the Dark Ages (DA) and the epoch of reionization (EoR) \cite{Zaroubi2013}.
Thus, in Sect.3 we study the imprint of the NP deviations on the CMB spectrum on the Sunyaev-Zel'dovich (SZ) effect in both thermal and non-thermal plasmas and we show that the main modifications to the standard SZ effect arise mainly in the very low-$\nu$ range where a very specific spectral shape of the modified SZ effect can be observed. We will show that at $\nu \simlt 500$ MHz these spectral features allow to measure directly the plasma frequency $x_p$.  Then in Sect.4 we will explore the effect on NP deviations of the CMB spectrum on the temperature brightness change of the cosmological 21-cm line background emerging from the Dark Ages and Epoch of Reionization and we predict the brightness temperature modifications observable with the next generation radio telescopes. We finally discuss our results in Sect.5.\\ 
Throughout the paper, we use a flat, vacuum--dominated cosmological model with $\Omega_m = 0.308$, $\Omega_{\Lambda} = 0.692$ and $H_0 =67.8$ km s$^{-1}$ Mpc$^{-1}$ \cite{Ade2015}.

\section{The spectral distribution of the CMB}
The spectral distribution of a photon gas which is in equilibrium with matter (e.g rarefied gases) at temperature $T$ is given by the Planck distribution which can be written in terms of the a-dimensional frequency $x=h \nu/k_B T$ as
\begin{equation}
\displaystyle I_0(x)=2 \frac{(k_B T)^3}{(h c)^2} \frac{x^3}{e^x-1} \; .
\label{planckian}
\end{equation}
For the case of the CMB, $T=T_{\text{CMB}}=2.725 \pm0.001$ K \cite{Fixsen2002} and $h$ is the Planck constant, $k_B$ is the Boltzmann constant and $c$ is the speed of light. In a ionized plasma with electron plasma frequency $x_p$, deviations from Planck distribution in eq.(\ref{planckian})  are possible because photons are suppressed at frequencies lower than $x_p$. By taking in consideration this effect, we can write the generalized Planck distribution for the CMB photons as follows \cite{Triger2010}
\begin{equation}
\displaystyle \tilde{I}_0(x;x_p)=2 \frac{(k_B T)^3}{(h c)^2} \frac{x^2}{e^x-1} \sqrt{x^2-x_p^2}\ \text{H}(x-x_p),
\label{nonplanckian}
\end{equation}
where $\text{H}(x-x_p)$ is the Heaviside step function (with values 1 for $x > x_p$ and 0 for $x \leq x_p$). The step-function $\text{H}(x-x_p)$ takes into account the fact that photons are suppressed at frequencies below $x_p$. The difference between the two distributions is more pronounced if $x_p$ is larger than 1 but nonetheless there are appreciable differences even for values of $x_p < 1$. The largest contrast between the generalized Planck distribution in eq.(\ref{nonplanckian}) and the standard Planck distribution in eq.(\ref{planckian}) is that the former has a cut-off frequency which depends on $x_p$. 
We also notice that the frequency at which the maximum of the CMB spectrum occurs shifts to higher frequencies as the cuff-frequency $x_p$ increases \cite{Triger2010} and that the CMB peak intensity decreases as the value of $x_p$ increases.
\begin{figure}
\includegraphics[width=140mm,height=110mm]{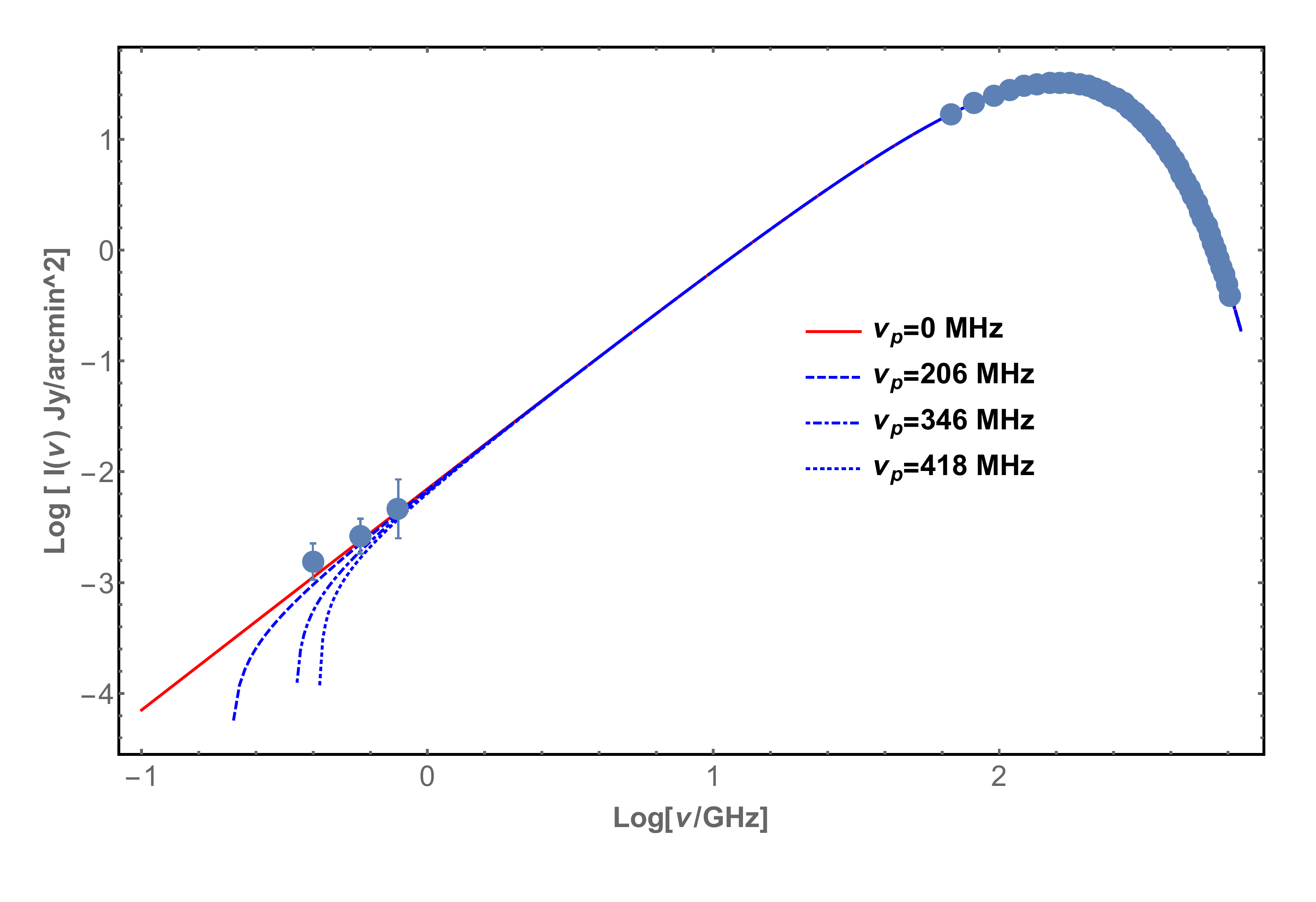}
\caption{The non-Planckian spectral distribution of the CMB  for different values of the plasma frequency $\nu_p$ as derived from the fit to the data. The experimental data shown here are from \cite{Howell1967,Sironi1990,Sironi1991} and COBE \cite{Fixsen1996}. Other data in the range $\sim 1.3-50$ GHz obtained from ground-based, ballon-borne and from the COBE-DR experiment (see http://asd.gsfc.nasa.gov/archive/arcade/cmb\_spectrum.html) are not shown here but they lie almost exactly along the curves of the CMB spectrum shown in the plot.}
\label{CMBfull}
\end{figure}
\begin{figure}
\includegraphics[width=120mm,height=100mm]{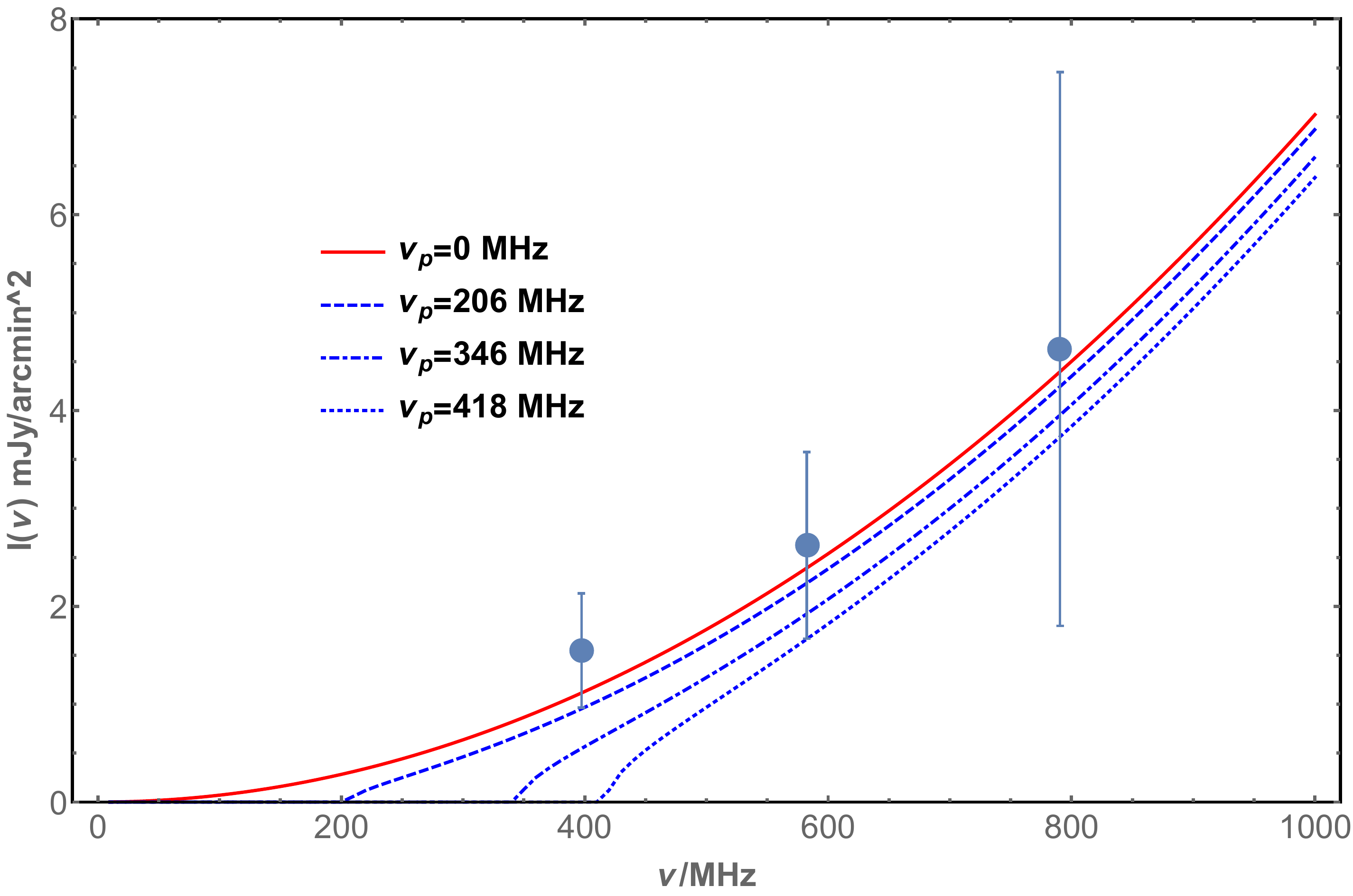}
\caption{The non-Planckian effects due to a finite value of the plasma frequency $\nu_p$ on the CMB spectrum at $\nu < 1$ GHz for different values of the upper limit on $\nu_p$ as obtained from the fit to the CMB spectrum data. Experimental data are from \cite{Howell1967,Sironi1990,Sironi1991}.}
\label{CMBlow}
\end{figure}

\begin{figure}
\begin{tabular}{c}
\includegraphics[width=80mm,height=60mm]{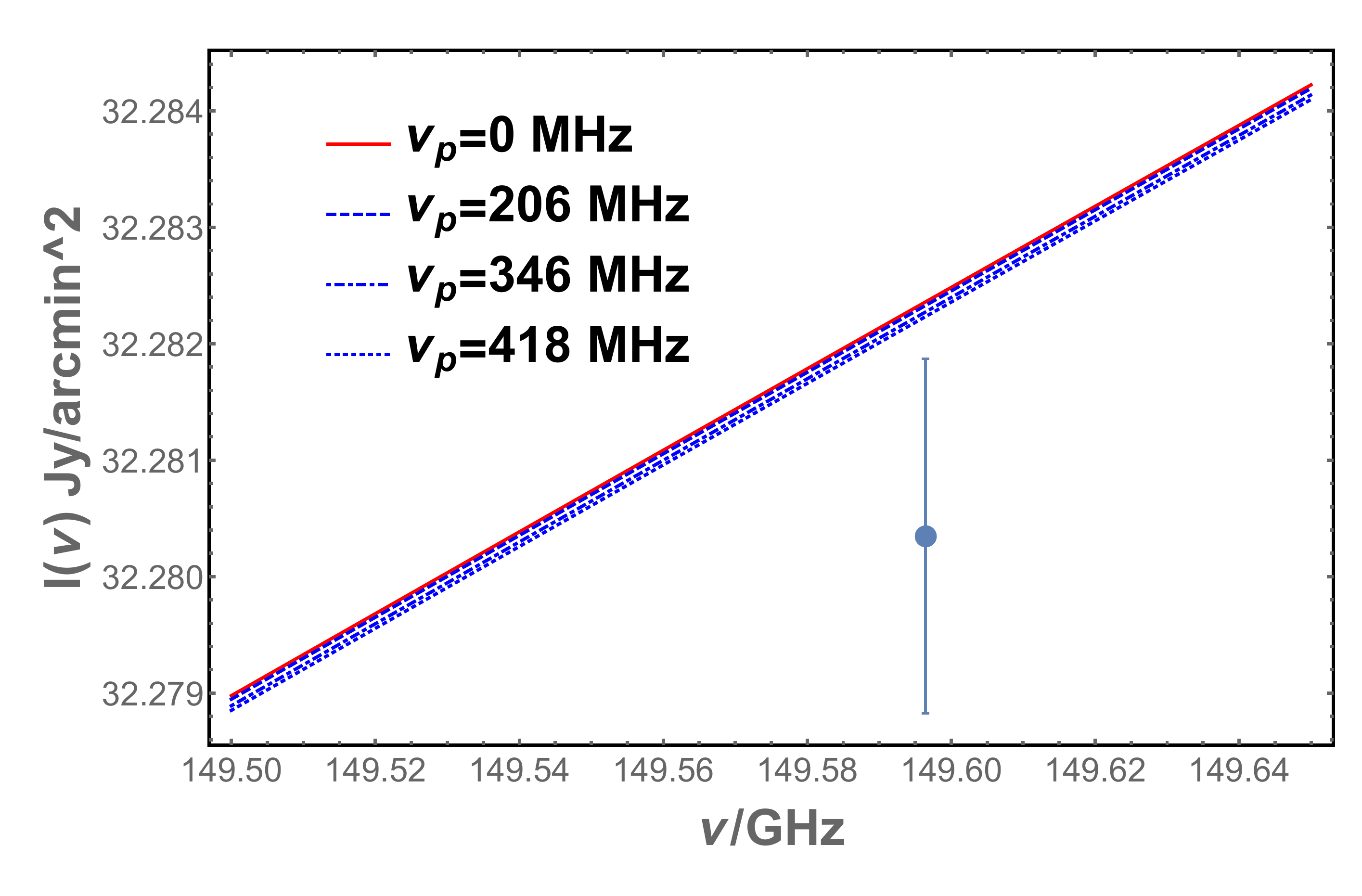}\\
\includegraphics[width=80mm,height=60mm]{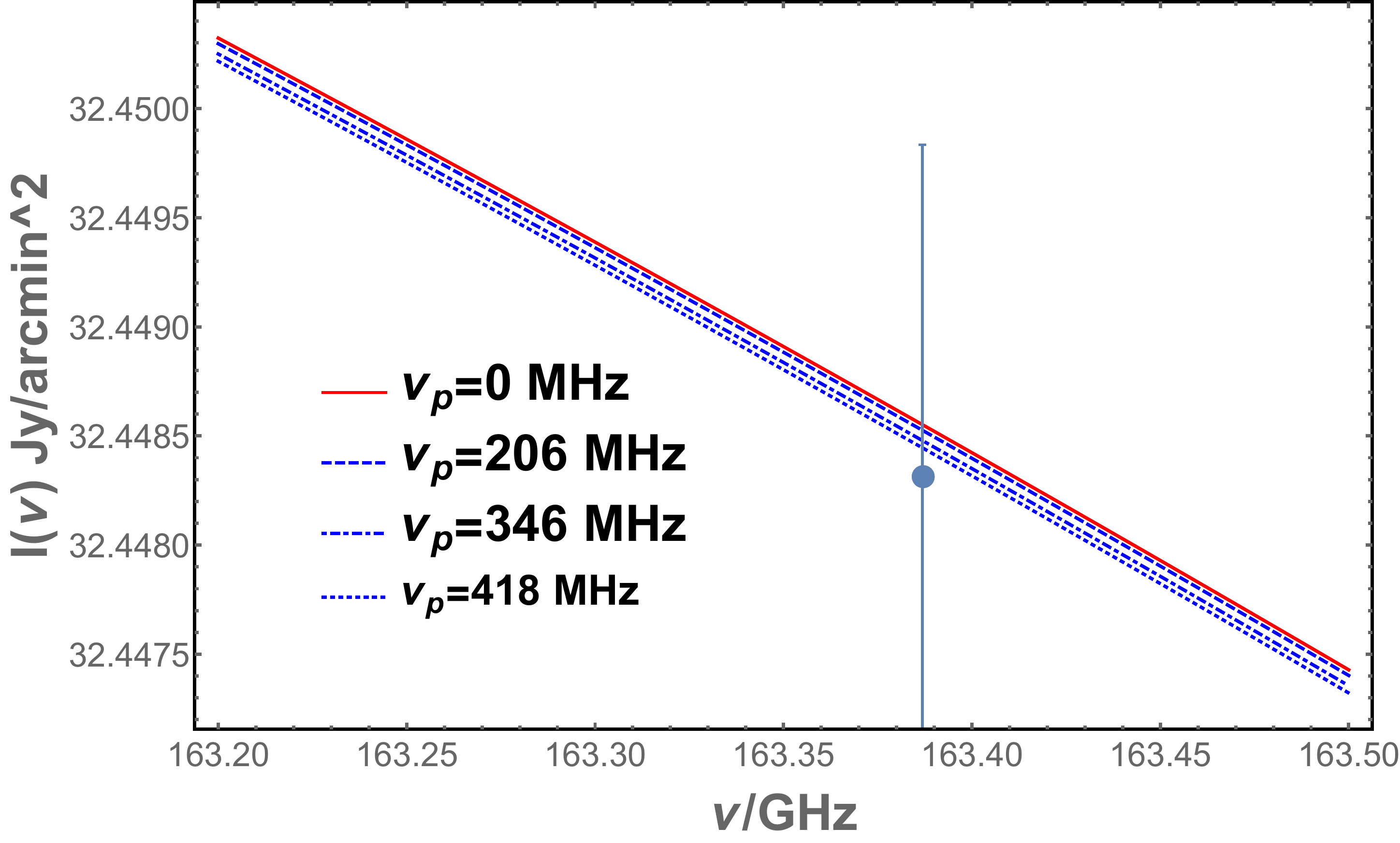}\\
\includegraphics[width=80mm,height=60mm]{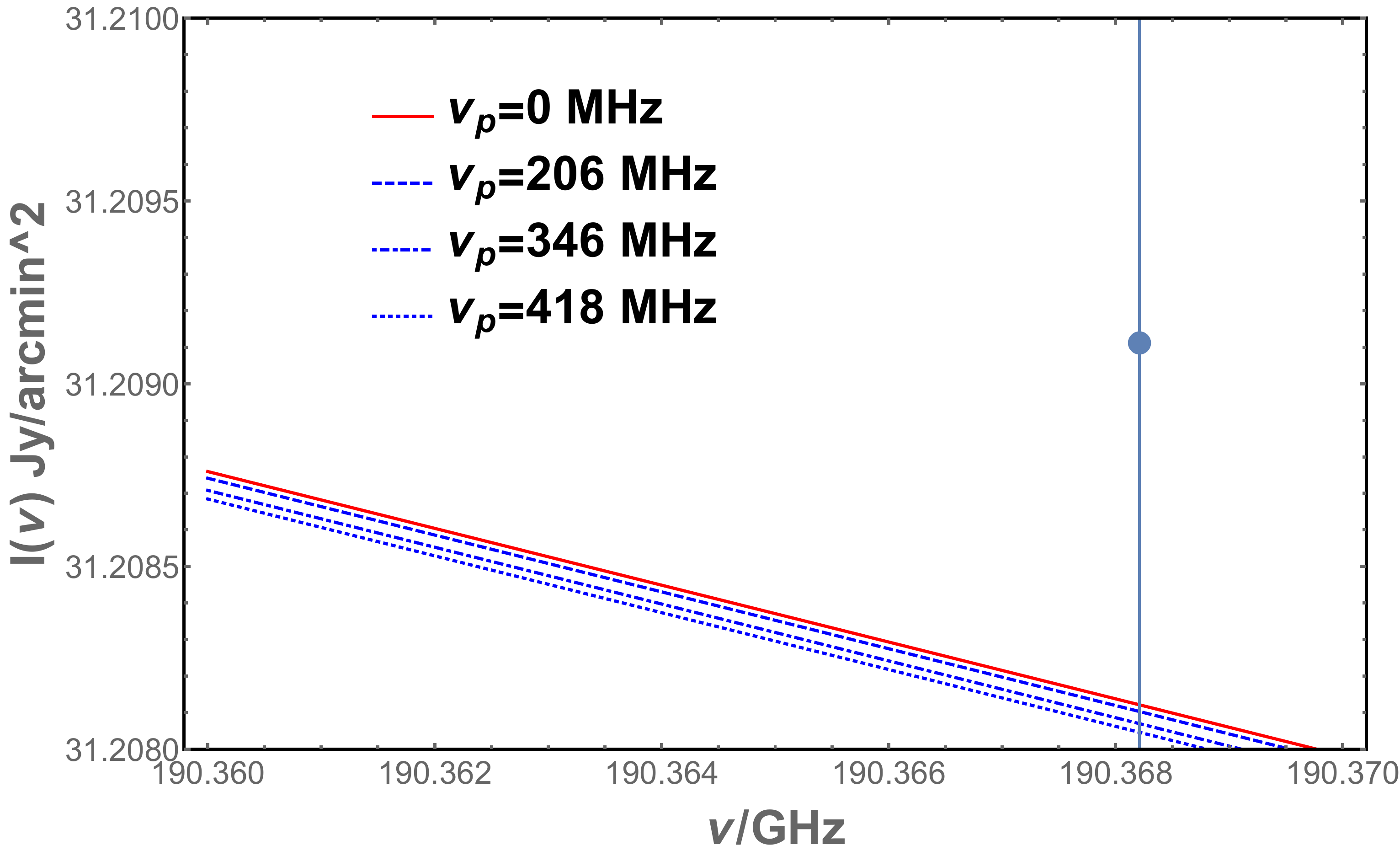}
\end{tabular}
\caption{The effect of non-Planckian distribution on the CMB spectrum in  frequency range 143 GHz to 191 GHz is shown for different values of the plasma frequency $x_p$ as derived in Sect.2. Experimental data are from COBE \cite{Fixsen1996}.}
\label{CMBpeak}
\end{figure}

In order to investigate the presence of non-Planckian effect on the CMB spectrum, we fit the NP distribution of the CMB spectrum in eq.(\ref{nonplanckian}) to the CMB data \cite{Fixsen1996},\cite{Howell1967,Sironi1990,Sironi1991} and we minimize the $\chi^2$ with respect to the plasma frequency $x_p$; in principle one can also use the CMB temperature $T$ as a free parameter: however, we set its value to the one measured by COBE-FIRAS that is very accurate with uncertainty of order of $\approx 0.03 \%$ and it is also determined in the high-$\nu$ range of the CMB spectrum (around its maximum) where the NP effects are small. With this fitting procedure we obtain a value of $x_p=0$ for the minimum $\chi^2$. This means that a pure Planckian spectrum is best fitting  the present data. However, the uncertainties in the data (in particular at low frequencies $\nu < 1$ GHz) allow us to set upper limits on the quantity $x_p$ which are $x_p=(3.63, 6.10, 7.36)\times10^{-3}$ for $1\sigma$, $2\sigma$ and $3\sigma$ confidence level (c.l.), corresponding to frequencies of 206, 346 and 418 MHz, respectively. 
In Fig \ref{CMBfull} we show the NP effects on the CMB spectrum over the full frequency range. 
This result indicates that it is not possible at the 3$\sigma$ c.l. to distinguish between the Planck spectrum and the NP modified spectrum based on the current knowledge of the CMB spectrum.  This fact opens hence an experimental window for the exploration of NP effects in the early universe.
We then show in Fig \ref{CMBlow} the CMB spectrum and its experimental uncertainties zooming around the low frequency region at $\nu <1$ GHz showing the modified CMB spectrum with the upper limits on $x_p$ at $1\sigma$, $2\sigma$ and $3\sigma$ c.l. 
These results show that the intensity of the CMB goes to zero below $\approx 206$ MHz, i.e. for the value of $x_p = 3.63 \times 10^{-3}$  at $1\sigma$ c.l., or below  $\approx 418$ MHz, i.e. for the value of $x_p = 7.36 \times 10^{-3}$  at $3\sigma$ c.l. In Fig \ref{CMBpeak}, we show the relative CMB spectrum variations around the peak of the CMB where the COBE  data \cite{Fixsen1996} are available (for the sake of illustration of the amplitude of the NP effects we zoom within the region 149--191 GHz around the maximum of the CMB spectrum).
As shown in Fig \ref{CMBpeak}, the CMB spectrum intensity decreases with increasing values of  $x_p$ and the frequency location of its maximum increases with increasing value of $x_p$. The amplitude of the CMB intensity variation due to NP effects is of order of $\approx$ mJy/arcmin$^2$ at low-$\nu$ while it is of order of $\approx 0.1$ mJy/arcmin$^2$ at high-$\nu$. Therefore, the best frequency region to look experimentally for these effects is at low radio frequencies, i.e. at $\nu \simlt 1000$ MHz.
Our analysis shows that precise measurements of the CMB spectrum at low radio frequency have the power to set strong constraints to NP effects due to the existence of a finite value of the plasma frequency at recombination and that one can probe these effects with the next generation experiments like the SKA (SKA-LOW  and SKA-MID \cite{Dewdneyetal2012}) and HERA \cite{Hera}.\\ 
Even though the CMB spectrum can be considered as a direct probe of NP effects in the early universe, its measurement at very low radio frequencies $\simlt 400$ MHz is challenging and it is subject to various experimental systematics and biases, and on the ability to subtract carefully radio foregrounds \cite{Oliveira2008}. The next generation low-$\nu$ radio telescopes (SKA, HERA) will increase our knowledge of the CMB spectrum in this frequency range, but the problem of foreground contamination and of the component separation of such diffuse signals at low frequency still remains.
Therefore it is interesting to explore the use of other observable quantities that depend directly on the CMB spectrum, like, e.g., the modified SZ effect \cite{ColafrancescoMarchegiani2014a} and the modifications to the 21-cm brightness temperature change w.r.t. the CMB. 
In the following Sections we will discuss these two additional probes of NP effects in the early universe.

\section{The modified Sunyaev Zel'dovich effect}
Another method which can be used to probe the existence of non-Planckian effect in the CMB is to use the modified SZ effect (see \cite{ColafrancescoMarchegiani2014a}, and \cite{Birkinshaw1999,Colafrancescoetal2003} for the general derivation of the SZ effect). The SZ effect is a spectral distortion of the CMB spectrum where the CMB photons interact with the electrons (thermal and non-thermal) residing in the atmosphere of galaxy clusters and other structures on large scales, like e.g. radio galaxies and galaxy halos. 
The SZ effect is a differential measure of the CMB spectrum on and off an area of the sky containing the selected cosmic structure and hence it is not affected by large-scale foregrounds in the observations at low-frequency (see \cite{Colafrancescoetal2015} for a discussion). 

The spectral distortion due to SZ effect on the CMB spectrum can be computed, at first order in the optical depth $\tau_e = \sigma_T \int d \ell n_e$,  as follows \cite{Colafrancescoetal2003}:
\begin{equation}
\displaystyle \Delta{I}(x)=\tau_e\bigg[\int ds P_1 (s){I}_0(x e^{-s}) - {I}_0(x)\bigg],
\label{szeplanck}
\end{equation}
where $P_1(s)$ is the single scattering redistribution function, obtained by summing the photon redistribution function $P_s(s,p)$, that gives the probability to have a logarithmic shift  $s=\ln (\nu'/\nu)$ in the photon frequency due to the ICS process by an electron with a-dimensional momentum $p=\beta \gamma$, over the electron momentum distribution function $f_e (p)$, normalized as to have $\int_0^\infty f_e(p) dp=1$ (see \cite{Ensslin2000,Colafrancescoetal2003}). 
This is written as
\begin{equation}
\displaystyle P_1(s)=\int_0^\infty dp f_e (p) P_s (s,p).
\label{p1s}
\end{equation}
For the calculation of the standard SZ effect, the quantity $I_0(x)$ is a Planck spectrum in the case of the CMB. Since the SZ effect is related directly to the spectrum of the CMB (see eq.\ref{szeplanck}), the existence of non-Planckian effect in the CMB background will be reflected directly in the SZ spectral distortion, and it is expected to be visible in particular around the cut-off frequency $x_p$. In order to compute the SZ effect for the case of a NP spectrum, eq.(\ref{nonplanckian}) is inserted into eq.(\ref{szeplanck}) instead of $I_0(x)$ as follows
\begin{equation}
\displaystyle \Delta \tilde{I}(x;x_p)=\tau_e\bigg[\int ds P_1 (s) \tilde{I}_0(x e^{-s};x_p) - \tilde{I}_0(x;x_p)\bigg].
\label{eq.deltai}
\end{equation}
One can notice directly from this equation that the final SZ spectral distortion in the case of NP plasma effect depends directly on $x_p$, hence on the plasma frequency $\nu_p$.

\begin{figure}
\begin{center}
\begin{tabular}{c c}
\includegraphics[width=80mm,height=75mm]{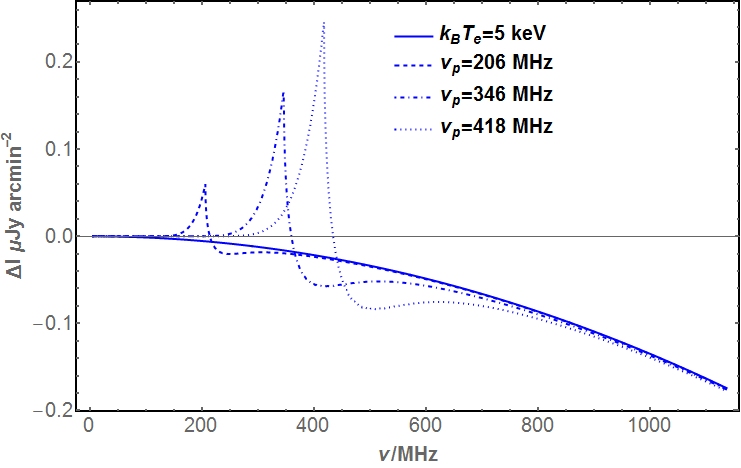} & \includegraphics[width=80mm,height=75mm]{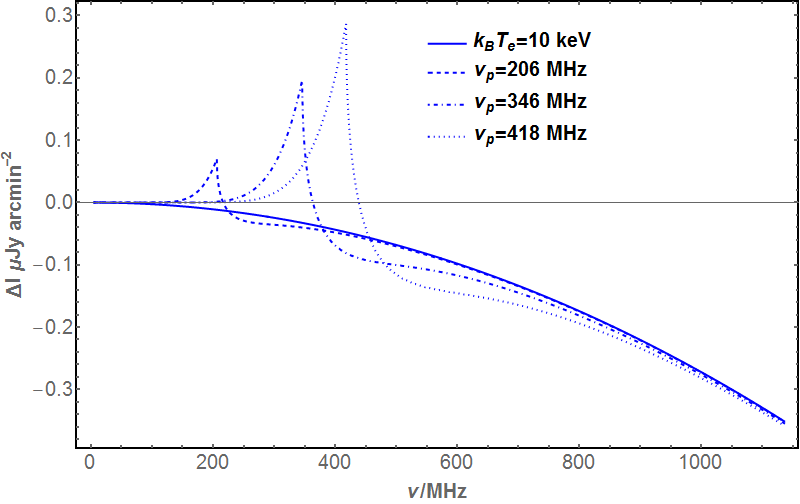}\\
\includegraphics[width=80mm,height=75mm]{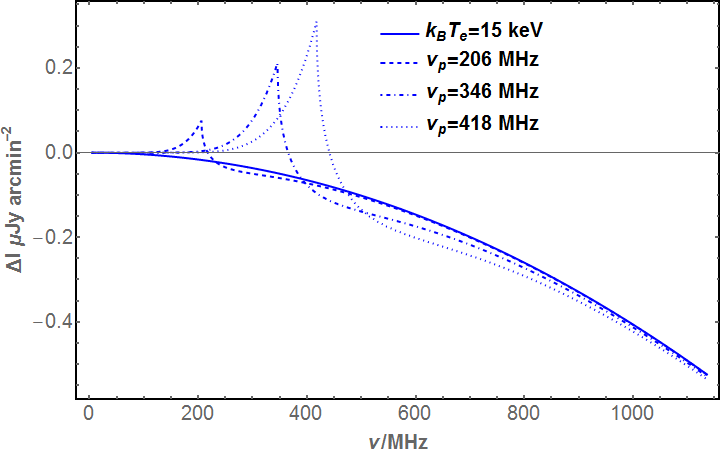} & \includegraphics[width=80mm,height=75mm]{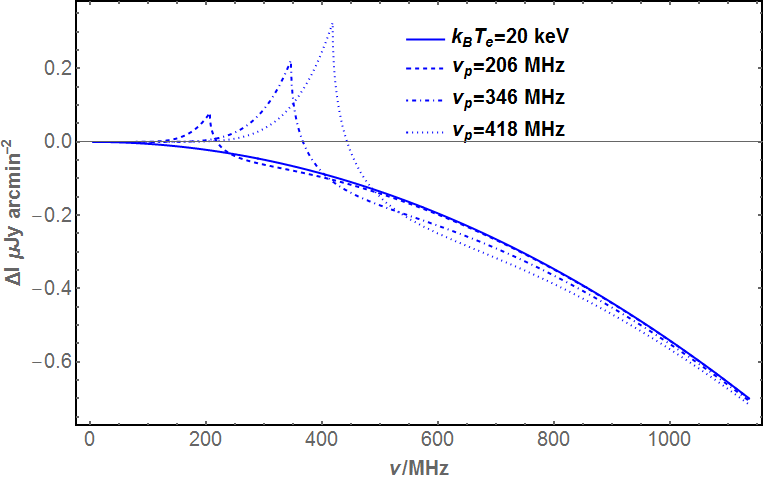}
\end{tabular}
\caption{The thermal SZ  effect spectral distortions computed for galaxy clusters with increasing plasma temperature (see various panels) for a standard Planck distribution (solid line) and including the effect of non-Planckian distribution of photons for the values of the plasma frequency $x_p$ derived at $1\sigma$, $2\sigma$ and $3\sigma$ level.  The cluster plasma optical depth is fixed to the value $\tau = 0.001$.}
\label{thermalsze}
\end{center}
\end{figure}

In the context of the SZ effect modified by NP effects (that we refer here to as SZ$_{NP}$) we first compute the thermal spectral distortion for an electron plasma within galaxy clusters with $\tau_e =10^{-3}$ and with different electron temperatures of 5 keV, 10 keV, 15 keV and 20 keV. We show in Fig \ref{thermalsze} the SZ$_{NP}$ spectrum for different values of  $x_p$ derived from the analysis of the CMB spectrum in the previous Sect.2. 
%
%
\begin{figure}
\begin{tabular}{c c}
\centering
\includegraphics[width=80mm,height=77mm]{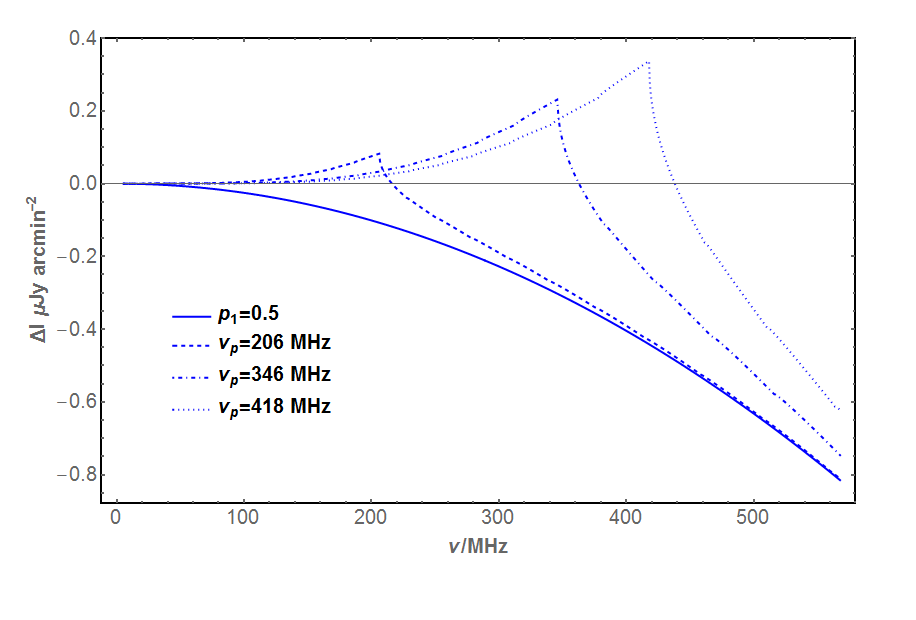}&\includegraphics[width=80mm,height=75mm]{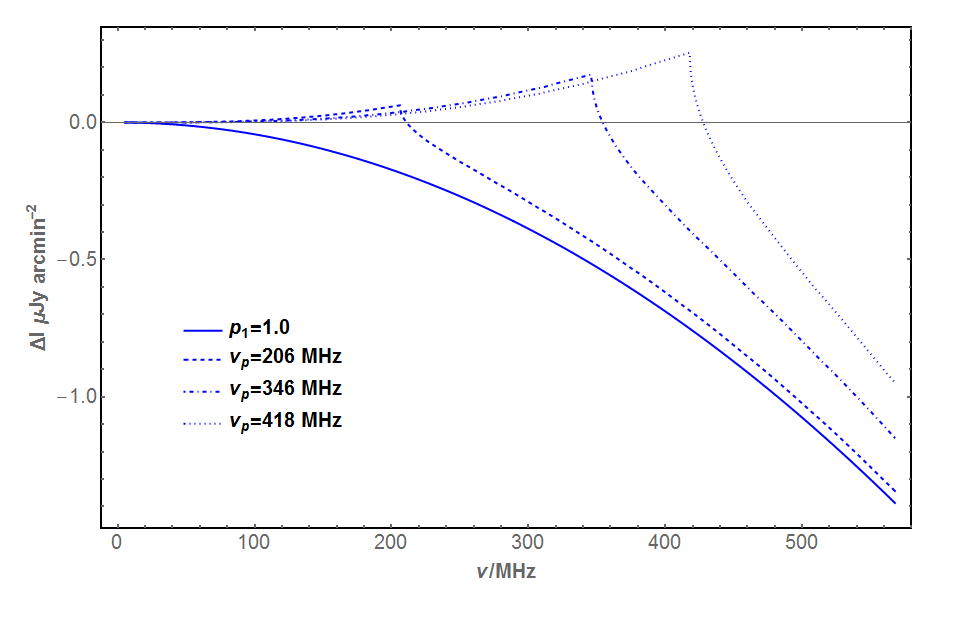}\\
\includegraphics[width=85mm,height=80mm]{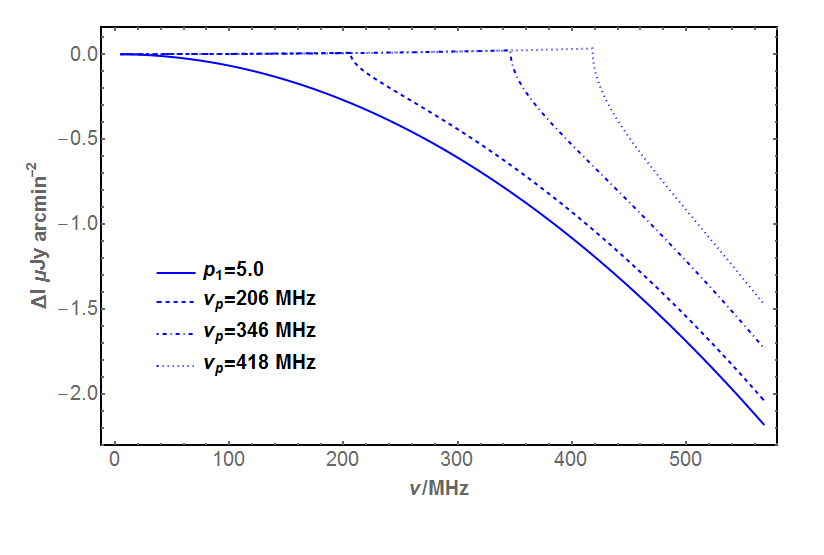}&\includegraphics[width=80mm,height=75mm]{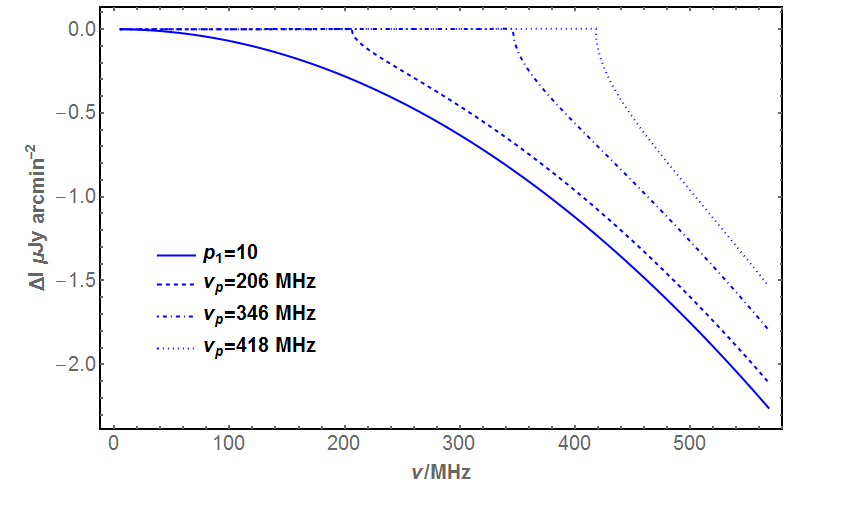}
\end{tabular}
\caption{The non-thermal SZ spectral distortion for the case of non-thermal plasmas  with increasing electron minimum momentum $p_1$  (see various panels) for a usual Planck distribution (solid line) of photons and the effect of non-Planckian distribution at $1\sigma$, $2\sigma$ and $3\sigma$ level computed in the case of a single power law for different minimum momentum $p_1$. We use here an optical depth of $\tau =0.001$.}
\label{nonthermalsze}
\end{figure}

We notice that the frequency at which the maxima of the SZ$_{NP}$ occurs is exactly $x_p$ and is independent of the cluster temperature. 
This spectral feature offers therefore a direct experimental probe of the plasma frequency in the early universe as  measured by the SZ$_{NP}$ in structures of the local universe. We stress that this is a unique characteristic of the SZ$_{NP}$ and allows to use observations of the local universe to infer directly the physics of the early universe.\\
We also show that the amplitude of the peak of SZ$_{NP}$ increases with the temperature and density (i.e. with the total Compton parameter $Y\propto \int d \ell P_e$, where $P_e$ is the total electron pressure) of the electron gas. 
The spectral shape of the thermal SZ$_{NP}$ is very peculiar and it is peaked in a quite narrow frequency range of the order of 
$\Delta \nu \approx 20$ MHz for $kT= 5$ keV and  $\Delta \nu \approx 50$ MHz for $kT= 20$ keV, reflecting hence the relativistic effects of the electron distribution in the photon re-distribution function $P_1(s)$. At frequencies larger than 1 GHz, the SZ$_{NP}$ becomes close to the standard Planckian one, i.e. the standard SZ effect calculated with the non-distorted CMB Planck spectrum \cite{Colafrancescoetal2003} as an input. 

The SZ$_{NP}$ can appear in the form of both thermal and non-thermal effects depending on the nature of the electron population producing the ICS of the CMB photons (i.e. thermal electrons in clusters for the thermal SZ$_{NP}$ and non-thermal or relativistic electrons for the non-thermal SZ$_{NP}$).
For completeness, therefore, we also compute the effect of non-planckian CMB distribution on the non-thermal SZ$_{NP}$ effect which is shown in Fig \ref{nonthermalsze} for a power law electron spectrum with spectral index $\alpha=2.5$ and
different values of the minimum momentum $p_1$ of the non-thermal electron distribution. In this case, the peak of the SZ$_{NP}$ decreases with increasing value of $p_1$ because the high-E electrons scatter photons to very high frequency.
In this case the peak of the SZ$_{NP}$ is less pronounced (see Fig.\ref{nonthermalsze}) but it is still located at the plasma frequency value $x_p$. The width of the SZ$_{NP}$  spectrum (see Fig.\ref{nonthermalsze}) is larger due to the enhanced impact of the relativistic effects of the high-E electron population and its shape reflects therefore the different nature of the scattering plasma. Possible observations of the SZ$_{NP}$ can therefore address the question of the intrinsic nature of the plasma in the target cosmic structure (e.g., galaxy clusters vs. radio galaxies).

\begin{figure}
\includegraphics[width=140mm,height=100mm]{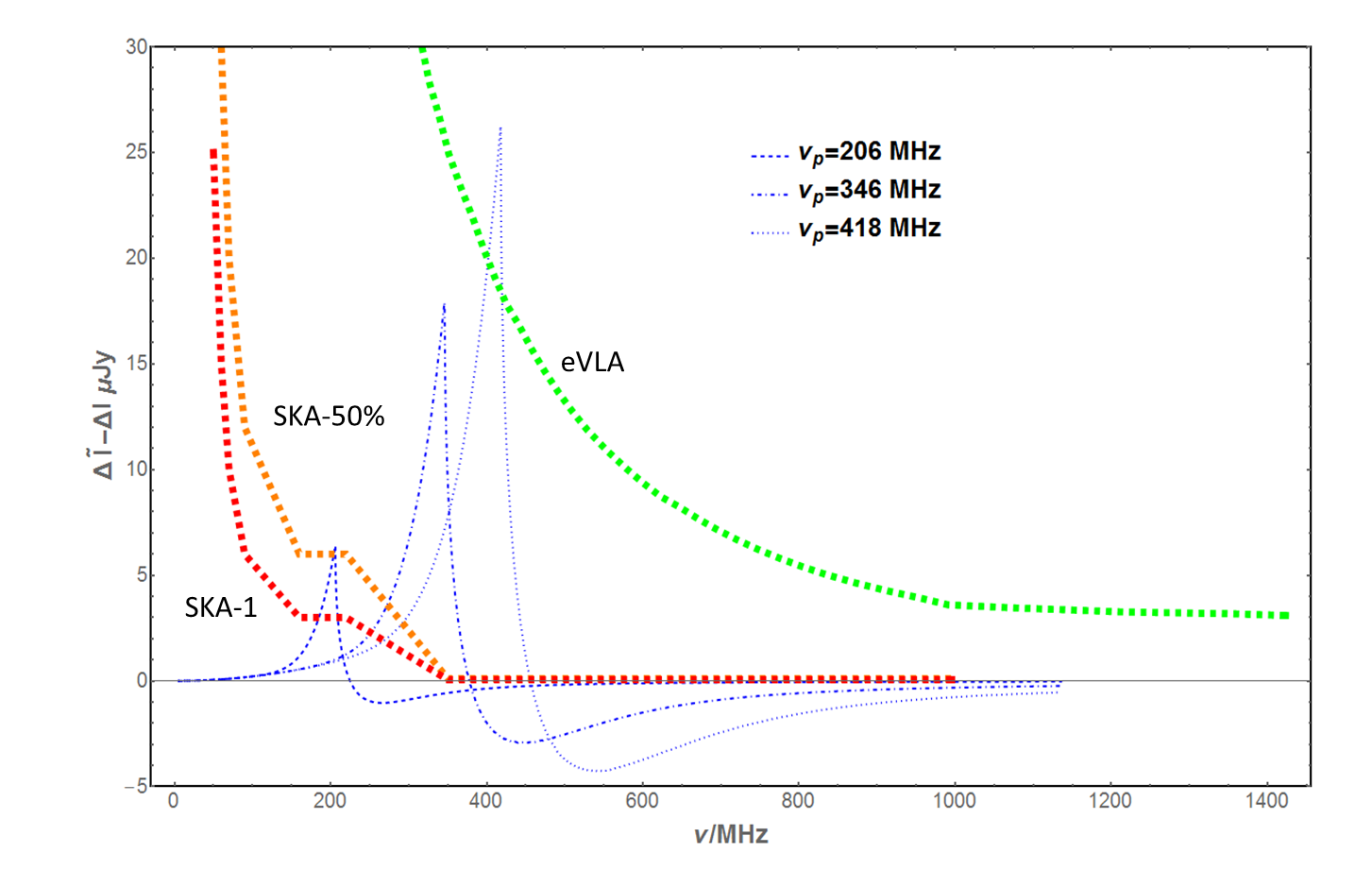}
\caption{The difference between the SZ$_{NP}$ and the Planckian SZ for a Bullet-like cluster with temperature of 15 keV (see \cite{bullettemp}) subtending an angle of $5^{\prime}$  to its $R_{500}$. An optical depth of $\tau =0.001$ is assumed. The SKA1 (red) and SKA-50\% (orange) sensitivity is calculated for 1000 hrs integration while the eVLA sensitivity (green) is calculated for 12 hrs integration.}
\label{skasze}
\end{figure}

In order to study the possible detectability of NP effect on the CMB, we have  compared the SZ$_{NP}$ spectrum integrated over an angle of $5^{\prime}$ radius of a Bullet-like cluster with temperature of $kT = 15$ keV. Fig \ref{skasze} shows the difference between the SZ spectrum $\Delta \tilde{I}$ and $\Delta I$ compared to the sensitivity of the SKA1 and SKA50\% for 1000 hrs of observations and to the eVLA sensitivity for 12 hrs of observation. We found that the values on the plasma frequency $\nu_p$ = 206, 346 and 418 MHz at 1, 2 and 3 $\sigma$ c.I, derived from the analysis of the CMB spectrum, can be detected with both SKA-LOW and SKA-MID band1 (350-1050 MHz).

\section{Modifications to the cosmological 21-cm background}
\begin{figure}[h!]
\includegraphics[scale=0.5,angle=180]{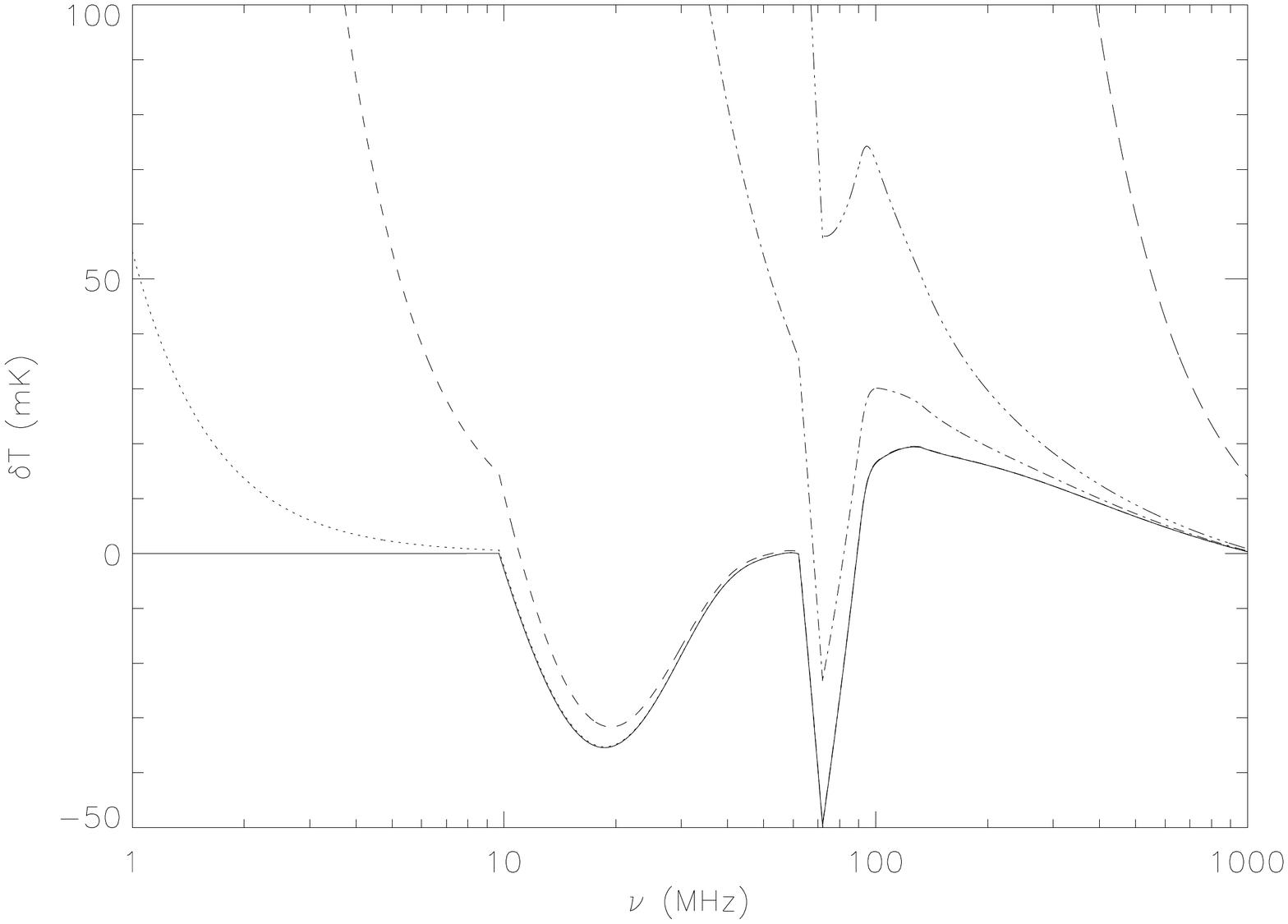}
\caption{The change in brigthness temperature $\delta \tilde{T}$ for $\nu_p=0$ (solid line), $\nu_p=0.2$ (dotted), 1 (dashed), 10 (dot-dashed), 20 (three dots-dashed) and 100 MHz (long dashed) as a function of the frequency.}
\label{dt21cm}
\end{figure}

The frequency range where NP effects can be observed (see Sects.2 and 3) is the same where the distortion of the CMB spectrum induced by the 21-cm background is more relevant \cite{Furlanetto2006}. For this reason we explore in this Section how NP effects on the CMB spectrum as in eq.(\ref{nonplanckian}) impact on the cosmological 21-cm background spectrum. To this aim, we discuss the results of our spectral analysis in term of the brightness temperature change given by
\begin{equation}
\delta T (\nu)= \frac{c^2}{2k_B \nu^2} \delta I(\nu).
\label{brigthness.temp}
\end{equation}
For a Planckian CMB spectrum, the 21-cm background appears as a perturbation on the CMB background, so we have:
\begin{equation}
\delta T(\nu) = T_{21}(\nu) - T_0 (\nu) ,
\end{equation}
where $T_{21}$ is the brightness temperature of the 21-cm background and $T_0$ is the brightness temperature of the CMB.
In the case of a non-planckian CMB spectrum, the previous relation can be generalized as follows
\begin{equation}
\delta \tilde{T}(\nu;\nu_p) = \tilde{T}_{21}(\nu;\nu_p) -\tilde{T_0} (\nu;\nu_p).
\end{equation}
Once we fixed the value of $x_p$, we can distinguish two frequencies regimes:\\
\textit{i)} $\nu<\nu_p$: in this case $\tilde{T_0}=0$, and therefore $\delta \tilde{T}=\tilde{T}_{21}$. If we assume that $\tilde{T}_{21} \approx T_{21}$, i.e. the 21-cm background is not sensibly changed by the NP spectral distortion, as it is expected because the main physical processes affecting the 21-cm background depend on the global temperature of the system which is not heavily affected by a distortion at small frequencies, then we can write:
\begin{equation}
\delta \tilde{T} = \delta T + T_0 ,
\end{equation}
so the frequency change of the brightness temperature is the same as in the Planckian case, but its amplitude is shifted by a value $T_0$ (that in the RJ region has a constant value equal to the CMB temperature);\\
\textit{ii)} $\nu>\nu_p$: in this case  we can write the NP modified spectrum in the RJ region as:
\begin{equation}
\tilde{I}_0 \propto x \sqrt{x^2-x_p^2} = x^2 \sqrt{1-\frac{x_p^2}{x^2}},
\end{equation}
and the corresponding brightness temperature is:
\begin{equation}
\tilde{T}_0= T_0 \sqrt{1-\frac{x_p^2}{x^2}}.
\end{equation}
By assuming again $\tilde{T}_{21} \approx T_{21}$, the change in the temperature brightness is given by:
\begin{equation}
\delta \tilde{T} = \delta T + T_0 - \tilde{T}_0 = \delta T + T_0 \left(1-\sqrt{1-\frac{x_p^2}{x^2}}\right).
\end{equation}
In Fig.\ref{dt21cm} we show the spectral change of $\delta \tilde{T}$ for different values of $\nu_p$ from 0.2 to 100 MHz, by assuming a standard benchmark model for $\delta T$ (as discussed in\cite{Cooray2006,ColafrancescoMarchegiani2014b}). 
For $\nu<\nu_p$ the frequency change of $\delta \tilde{T}$ is the same $\delta T$ obtained in the standard Planckian case but its amplitude is increased by the value $T_0$; for $\nu \simgt \nu_p$ the quantity $\delta \tilde{T}$ decreases very rapidly with increasing frequency, and for $\nu \gg  \nu_p$ it then becomes equal to the standard value $\delta T$ in the Planckian case.
Fig.\ref{dt21cm} shows that even low values of the plasma frequency $\nu_p \simlt 206$ MHz (i.e., the 1 $\sigma$ upper limit set by the CMB spectrum analysis in Sect.2) can largely change the 21-cm radiation background spectrum and hence our ability to recover  the history of the EoR and of the DA.
NP effects do not change the physical mechanisms that are working during these epochs but change the brightness temperature contrast $\delta \tilde{T}$ relative to the CMB  that we can use to prove these remote cosmic epochs.
Interestingly,  the brightness temperature contrast $\delta \tilde{T}$ in the case in which NP effects are present on the CMB is increased w.r.t. the standard assumption of a pure Planckian CMB spectrum. This fact on one hand yields larger signals that could be detected more easily even with precursors of the SKA and HERA, and on the other hand provides an effective possibility to set limits or even prove the existence of NP effects in the early universe by looking at the global spectrum of the 21-cm background.
Available limits from the PAPER experiment \cite{Parson2014} in the 100--200 MHz provide limits only on the average temperature brightness of $\langle T_b \rangle < 275$ and $291$ mK for values of the ionization fraction $x_i = 0.5$ and $x_i=0.75$ of the ionization power spectra, respectively, which cannot be unfortunately directly compared to the predictions for the quantity $\delta T$ used in our paper. 
The next generation SKA and HERA low-frequency interferometers  will have the possibility to improve the knowledge on the brightness temperature contrast $\delta T$ down to levels of $\sim 3$ mK ($\sim 0.3$ mK) with SKA phase-1 (SKA phase-2) at $3 \sigma$ c.l. in about 1000 hrs at 150 MHz \cite{Koopmans2010}  and possibly cover a wider frequency range, thus allowing to set much stronger limits on $\nu_p$ from  the spectrum of $\delta T (\nu)$.

The SKA1-LOW band 50--350 MHz will be able to set limits only on plasma frequency values $\nu_p \simgt 10 $ MHz because for lower values of $\nu_p$ of order of a few MHz the NP modified 21-cm background spectrum  is very close to the Planckian case at the lower end of the SKA-LOW band (see Fig.\ref{dt21cm}). Therefore, the SKA-LOW will have a sensitivity to the NP effects at the redshift (frequency) of the EoR. 
We speculate here that other sources of more recent NP effects can arise during the EoR due to the increasing ionization fraction of the cosmic plasma at this epoch caused by Ly-$\alpha$ interactions, heating due to UV/X-ray primordial sources and DM annihilation. The treatment of these secondary NP effects is well beyond the purposes of this initial paper and will be addressed more specifically elsewhere.

\section{Discussion and conclusions}

We have performed for the first time in this paper an  analysis of NP effects on the cosmological radio background and we derived predictions for their amplitude on three different observables: the CMB spectrum, the SZ effect and the 21-cm temperature brightness change.\\
We have shown that NP effect can manifest on the CMB spectrum at $\nu \simlt 400$ MHz as a drastic cut-off in the intensity of the CMB (see Fig.\ref{CMBlow}). Using the CMB data of COBE and lower frequency measurements, we derived  for the first time upper limits on the plasma frequency $\nu_p$ = 206, 346 and 418 MHz at 1, 2, 3 $\sigma$ c.l., respectively. 
The difference between the pure BB Planck spectrum and the one modified by NP effects at these low frequencies is of the order of mJy/arcmin$^2$ and this difference becomes smaller at higher frequencies ($\nu \approx 150$ GHz) where it is of the order of 0.1 mJy/arcmin$^2$, thus indicating that the experimental route to probe NP effects in the early universe is to observe the cosmological radio background at very low frequencies.

We have also computed the SZ$_{NP}$ effect using the upper limits on $\nu_p$ allowed by the present data of the CMB (Sect.2) and it has been found that the NP effect produces a unique imprint on the SZ$_{NP}$ effect spectrum at $\nu \simlt 1$ GHz,  i.e., a peak located exactly at the plasma frequency $\nu_p$ and that this is independent of the cluster parameters (such as its temperature or optical depth). This offers a way to measure directly and unambiguously the plasma frequency in the early universe at the epoch of recombination by using (in principle only one) cluster in the local universe, thus opening a unique window for the experimental exploration of plasma effects in the early universe. We have shown that the SKA-LOW and SKA-MID has the potential to observe such a signal integrating over the central regions ($\approx 5^{\prime}$ radius) of high-temperature ($k_B T \sim 15$ keV) clusters.\\
Observing the SZ$_{NP}$ also benefits from its differential nature thus being less affected by the large impact of large-scale foreground emission that is one of the main systematic biases that limits the study of the intensity spectrum of the cosmological background radiation. The SZ$_{NP}$ has also the appealing property that we can study the presence of NP effects in the early universe by looking at very local cosmic structures for which the structural parameters are known with high accuracy. Finally, we mention that studies of NP effects through the SZ$_{NP}$  can be done by intensive observations of only
one galaxy cluster, or with a stacked spectrum of a few well known clusters, thus avoiding the need of large statistical studies of source populations or wide area surveys.

Finally, we have shown that future low-frequency observations of the cosmological 21-cm brightness temperature spectral changes have the possibility to set global constraints on NP effects by constraining the spectral variations of $\delta \tilde{T}$ induced by the plasma frequency value at the epoch of recombination.
We discussed, in this context, that even moderate  limits on the average brightness temperature of the 21-cm background obtained with SKA precursors, like e.g. the PAPER experiment, are able to start limiting the possible values of $\nu_p$ in its high-frequency domain (of order of 100 to a few hundreds MHz).

In conclusion, we have demonstrated that the study of low-frequency cosmological radio background has a strong and unique potential for proving the physics of the early universe. This paper aligns such the idea of proving NP effects with our previous studies of the photon decay effects in the early universe \cite{ColafrancescoMarchegiani2014a} and on the study of the DA and EoR through the SZE-21cm \cite{ColafrancescoMarchegiani2014b,Colafrancescoetal2015}, and indicates that this area of investigation of the fundamental physics of the universe will receive a boost with the next generation high-sensitivity radio telescopes like the SKA.

\section*{Acknowledgments}

S.C. acknowledges support by the South African Research Chairs Initiative of the Department of Science and Technology and National Research Foundation and by the Square Kilometre Array (SKA). P.M. and M.S.E. acknowledge support from the DST/NRF SKA post-graduate bursary initiative.

\end{document}